\documentclass[twocolumn,aps,showpacs]{revtex4}
\usepackage{bm}
\usepackage{amsmath}
\usepackage{amssymb}
\usepackage{psfrag}
\begin{document}
%-------------------------------------------------------------------------
\newcommand{\Painleve}{Painlev\'e}
%-------------------------------------------------------------------------
%\title{Hydrodynamics in General Relativistic form}
%-------------------------------------------------------------------------
%\title{General Relativity in Hydrodynamic Form}
%-------------------------------------------------------------------------
%\title{Curvature of space-time in \Painleve-Gullstrand effective geometry}
%-------------------------------------------------------------------------
\title{On the space-time curvature experienced by quasiparticle excitations
in the \Painleve-Gullstrand effective  geometry} 
%-------------------------------------------------------------------------
\author{Uwe R. Fischer}
\affiliation{Department of Physics, 
University of Illinois at Urbana-Champaign\\
1110 West Green Street, Urbana, Illinois 61801--3080}  
%-------------------------------------------------------------------------
\author{Matt Visser}
\affiliation{School of Mathematical and Computing Sciences \\
Victoria University of Wellington 
\\ P.O. Box 600, Wellington, New Zealand}
%\affiliation{Physics Department, Washington University\\  
%Saint Louis, Missouri 63130--4899\\
%\ \ \ } %used to subvert defective spacing conventions in revtex4
%-------------------------------------------------------------------------
%-------------------------------------------------------------------------
\begin{abstract}
%-------------------------------------------------------------------------
We consider quasiparticle propagation in constant-speed-of-sound
(iso-tachic) and almost incompressible (iso-pycnal) hydrodynamic
flows, using the technical machinery of general relativity to
investigate the ``effective space-time geometry'' that is probed by
the quasiparticles. This effective geometry, described for the
quasiparticles of condensed matter systems by the
{\Painleve}--Gullstrand metric, generally exhibits curvature (in the
sense of Riemann), and many features of quasiparticle propagation can
be re-phrased in terms of null geodesics, Killing vectors, and Jacobi
fields. As particular examples of hydrodynamic flow we consider shear
flow, a constant-circulation vortex, flow past an impenetrable
cylinder, and rigid rotation.

%-------------------------------------------------------------------------
\end{abstract}
%-------------------------------------------------------------------------
%-------------------------------------------------------------------------
%\date{22 March 2002; file: {\sf gr-hydro.tex}; version 0.03; \LaTeX-ed \today}
%-------------------------------------------------------------------------
\maketitle
%-------------------------------------------------------------------------
% Local definitions
%-------------------------------------------------------------------------
%-------------------------------------------------------------------------
\newcommand{\MF}{{\large{\manual META}\-{\manual FONT}}}
\newcommand{\manual}{rm}        % Substitute rm (Roman) font.
\newcommand\bs{\char '134 }     % add backslash char to \tt font %
\newcommand{\kbar}{{{\bf --}}\hspace{-5.5pt}$\kappa$}
%-------------------------------------------------------------------------
\newcommand{\be}{\begin{equation}}
\newcommand{\ee}{\end{equation}}
\newcommand{\bea}{\begin{eqnarray}}
\newcommand{\eea}{\end{eqnarray}}
%-------------------------------------------------------------------------
\def\ii{{\hat\imath}}
\def\jj{{\hat\jmath}}
\def\kk{{\hat k}}
\def\lll{{\hat l}} 
\def\tt{{\hat t}}
\def\xx{{\hat x}}
\def\yy{{\hat y}}
\def\zz{{\hat z}}
%--------------------------------------------------------------------------
\def\Iordanskii{Iordanski\v\i}
\def\projection{\hbox{projection}}
\def\d{{\mathrm{d}}}
\def\ie{{\emph{i.e.}}}
\def\etc{{\emph{etc.}}}
%-------------------------------------------------------------------------
\section{Introduction} 
%-------------------------------------------------------------------------
The description of many natural phenomena is most vividly carried out
in terms of hydrodynamics, because the concept of a streaming liquid
elucidates and helps to understand the physical significance and
structure of an underlying theory \cite{madelung}.  In its classical
sense \cite{lamb,milne-t}, hydrodynamics describes the motion of a
continuum, characterized by a velocity and density distribution, which
for a perfect fluid and in the nonrelativistic limit is described by
the Euler and continuity equations.  It has been recognized about
twenty years ago by Unruh \cite{unruh}, that the propagation of small
perturbations on such a hydrodynamic background, which is itself
governed by a continuum version of Newtonian physics, may be cast into
the form of a ``relativistic'' scalar wave equation
\begin{equation}
\square\,  \Phi\equiv 
\frac1{\sqrt{-g}}\partial_\mu 
\left(\sqrt{-g} \; g^{\mu\nu} \; \partial_\nu \Phi\right) =0
\label{E:dalembertian}
\end{equation} 
for the velocity potential $\Phi$ of the perturbations.  The
disturbances propagate in an effective space-time with metric
$g_{\mu\nu}$, which is in general curved.  The metric $g_{\mu\nu}$ was
later on shown to be of the {\Painleve}--Gullstrand form \cite{PG},
originally invented as an alternative to the Schwarzschild form of the
solution of the Einstein equations for a point mass source.  With the
advent of effective curved space-time theories, it became apparent
that the \Painleve--Gullstrand representation of the metric appears in
a host of such theories.  They comprise, besides the conventional
Euler fluid \cite{unruh,vissersonic}, superfluid $^3\!$He-A
\cite{GrishaPhysicsReports,GrishaPGBH}, atomic Bose-condensed vapors
\cite{BoseCondensate,CSM}, and general dielectric (quantum)
matter~\cite{Gordon1923,UlfPRA2000,SchuetzholdDielectricBH}.  

An interesting and important feature of the \Painleve--Gullstrand
metric is that it continues to give an appropriate physical
description for quasiparticle propagation even when the effective
space-time possesses a horizon~\cite{LandauHorizon}. This occurs
because the condensed matter origin of the metric in the
{\Painleve}--Gullstrand form is the spectrum of elementary excitations
(quasiparticles) \cite{Lieb1963}, which is {\em primary}.  This
physical energy spectrum, from which the metric is {\em obtained}
using the fact that for massless quasiparticles the energy spectrum is
\be 
g^{\mu\nu}p_\mu p_\nu=0, \label{pquadrat}
\ee 
which must be well-defined and, in particular, real everywhere in the
system.  In contrast, for the Schwarzschild form of the metric the
spectrum reads
\begin{equation}
E^2=
c^2\left(1-{r_{\rm S}\over r}\right)^2 p_r^2+
c^2\left(1-{r_{\rm S}\over r}\right) p_\perp^2~,
\label{SpectrumInSchwarzschild}
\end{equation}
where $r_{\rm S}$ is the usual Schwarzschild radius and $p_r,p_\perp$
are radial and transverse components of the
quasiparticle momentum, respectively.  The velocity $c$ plays the role
of the speed of light and is equal to the sound speed for phonons.
This ``Schwarzschild form'' of the spectrum exhibits imaginary mode
frequencies and consequently leads to instability of the condensed
matter system if a horizon is present, because it has sections of the
transverse momentum $p_\perp$ which result in $E^2<0$ inside the
horizon.  The {\Painleve}--Gullstrand metric, on the other hand, gives
real frequencies throughout a condensed matter system possessing a
quasiparticle horizon, which can thus be stable.

The non-equivalence of Schwarzschild and \Painleve--Gullstrand form of
the metric is related to the fact that the coordinate transformation
relating the Schwarzschild solution and the \Painleve--Gullstrand
representation becomes singular at the horizon \cite{LandauHorizon}.
This fact has, {\it inter alia}, led to the usage of
\Painleve--Gullstrand co-ordinates for investigations of Hawking
radiation in the ``conventional'' black hole context of gravitational
theory~\cite{parikhwilczek,SchuetzholdBH}, because these co-ordinates
are {\em nonsingular} through the horizon, making the appropriate
vacuum definition there much simpler.

The intrinsic characteristics of a curved space-time are described in
a covariant way by the Riemann tensor \cite{Riemann,MTW}. Our
objective in this paper is to describe the Riemannian curvature of the
effective spaces described by the \Painleve--Gullstrand metric, in the
underlying hydrodynamic terms appropriate to a flowing background
fluid.  We shall focus on two physical situations: quasiparticles in
flows with a constant speed of sound (iso-tachic flows), and
quasiparticles in an almost incompressible (iso-pycnal) hydrodynamic
flow. By ``almost incompressible'' we mean that we take both the
background density and the quasiparticle propagation speed relative to
the medium to be constants, and concentrate on those effects that are
due to motion of the medium, \emph{i.e.}, its velocity distribution.
In other words, even if a fluid has a constant ``refractive index'',
focussing and defocussing effects can be engendered through motion of
the fluid.

As particularly interesting examples we demonstrate how the tracks of
quasiparticles are distorted by propagation through a shear flow, a
constant-circulation vortex flow, around an impenetrable cylinder, and
how they propagate through a rigidly rotating fluid. In a more general
context we provide a local definition of ``focal length'' in terms of
the Riemann tensor, and show how the affine and ``natural'' (using the
Newtonian background time) parameterizations of null geodesics can be
related to each other.

%-------------------------------------------------------------------------
\section{\Painleve--Gullstrand curvature in 3+1 dimensions}
%-------------------------------------------------------------------------

In the following discussion the quasiparticle spectrum is assumed to
be linear in the fluid rest frame for ``small'' quasiparticle momenta,
$E= c |{\bm p}|$ corresponding to (\ref{pquadrat}), and deviating from
linearity for momenta approaching the ``Planck scale'' of the system
at hand.  In general the (3+1)-dimensional \Painleve-Gullstrand
metric~\cite{PG} reads
\be
g_{tt}= -{\rho\over c}\,[c^2 -{\bm v}^2],
\quad g_{ti} = -{\rho\over c}\, v_i,
\quad g_{ij}={\rho\over c}\, \delta_{ij}.
\label{PGmetric3+1-true}
\ee
That is, the metric has space-time interval
\begin{equation}
\d s^2 = \frac{\rho}{c}
\left[-c^2\d t^2 + \delta_{ij} (\d x^i - v^i \d t) (\d x^j - v^j \d t)\right]. 
\label{lineelement}
\end{equation}
By special convention, the indices on the 3-velocity are always raised
and lowered using the flat 3-dimensional Cartesian metric so that $v_i
= v^i$.

In the case of irrotational fluid flow (for instance in a superfluid
outside the cores of the (singular) quantized vortices), the
d'Alembertian equation (\ref{E:dalembertian}) can be derived directly
from a linearization procedure based on the Euler and continuity
equations~\cite{unruh,vissersonic}; the existence and relevance of the
\Painleve--Gullstrand effective metric then follows as a rigorous
theorem. If distributed vorticity is present, the situation is more
subtle~\cite{stone-et-al}: In hydrodynamics with distributed vorticity
one obtains a rather complicated system of coupled differential
equations, one of which contains the d'Alembertian operator (and
therefore also contains the effective metric) as a subsidiary
quantity~\cite{stone-et-al}.  Thus for hydrodynamics with distributed
vorticity, the effective metric is not the whole story --- but
certainly an important part of the story. In particular, if one
appeals to the {\emph{eikonal}} approximation (in this context
identical to the WKB approximation) one can derive Pierce's
approximate wave equation~\cite{Pierce}. In this approximation one can
write down the quasiparticle spectrum directly in terms of the
effective metric~\cite{stone-et-al}.

Note that the constant-time hypersurfaces are conformal to ordinary
flat Cartesian space. As long as we are interested in quasiparticles
that propagate along the null cones of this effective metric (that is,
quasiparticles moving at the speed $c$ relative to the medium), it is
permissible to neglect the overall conformal factor of $\rho/c$ and
consider the simplified metric
\be
g_{tt}= -[c^2 -{\bm v}^2],
\qquad g_{ti} = -v_i,
\qquad g_{ij}=\delta_{ij}.
\label{PGmetric3+1-c}
\ee
(This is simply the statement that conformal transformations leave 
null curves and, in particular, null geodesics, invariant.)  The
inverse of this simplified metric is
\be
g^{tt}= -{1\over c^2} ,
\quad g^{ti}= -{v^i\over c^2} ,
\quad g^{ij}=  \delta^{ij}-{v^i v^j\over c^2}.  
\label{PGmetric3+1-inverse-c}
\ee
Note that the Newtonian time parameter $t$ provides a preferred
foliation of the spacetime into space+time, and that this preferred
foliation will prove very useful.

%-------------------------------------------------------------------------
%\subsection{Iso-tachic flows}
%-------------------------------------------------------------------------

Suppose now that the speed of sound is iso-tachic, independent of
position and time. Then we can choose coordinates to set the speed $c$
of linear quasiparticle dispersion equal to unity, a convention adopted
in the formulae below.  The
(3+1)-dimensional \Painleve-Gullstrand metric~\cite{PG} then reads
\be
g_{tt}= -1 +{\bm v}^2,\qquad g_{ti} = -v_i,\qquad g_{ij}=\delta_{ij}.
\label{PGmetric3+1}
\ee
In general relativistic language the lapse function in the ADM
formulation \cite{MTW} is now unity and all the spacetime curvature is
encoded in the shift function --- which here describes the physical
velocity of the fluid.  The inverse metric is
\be
g^{tt}= -1 ,\;\qquad g^{ti}= -v^i ,\qquad g^{ij}= 
\delta^{ij}-v^i v^j.  
\label{PGmetric3+1-inverse}
\ee

Turning to the computation of curvature, the 24 independent connection
coefficients read (cf. \cite{MikePhysRevE})
\begin{eqnarray}
\Gamma^t{}_{ij} &=& D_{ij}\,, 
\nonumber\\
\Gamma^t{}_{tt} &=& v_i v_k \; D_{ik} = \frac12 ({\bm v}\cdot \nabla) {\bm v}^2\,,
\nonumber\\ 
\Gamma^t{}_{ti} &=& -v_j \; D_{ij}\,,
\nonumber\\
\Gamma^i{}_{jk} &=& v_i \; D_{jk}\,, 
\nonumber\\
\Gamma^i{}_{tt} &=& -\partial_t v_i 
- v_k \;\partial_i v_k + v_i v_l v_k \; D_{lk}\,,
\nonumber\\
&=& -\partial_t v_i 
-  \frac12 \left(\delta^{ij} - v^i v^j \right)\;\partial_j  {\bm v}^2\,, 
\nonumber\\
\Gamma^i{}_{tj} &=& -v_i v_k \; D_{jk} +\Omega_{ij}\,.
\end{eqnarray}
Here we have defined the deformation rate and angular velocity tensors
by
\begin{eqnarray}
D_{ij} &=& \frac12 \left(\partial_i v_j + \partial_j v_i \right) \nonumber\\
&=& \partial_{(i} v_{j)} = D_{ji}\,,
\nonumber
\\
{\rm Tr}\, {\bm D} &=&   {\rm div}\, {\bm v} \,,
\nonumber\\
\Omega_{ij}&=& \frac12\left(\partial_i v_j - \partial_j v_i\right)\nonumber\\
&=& \partial_{[i} v_{j]}
= - \Omega_{ji}\,.
\end{eqnarray}
The deformation rate is in general relativistic language the extrinsic
curvature of the constant-time hypersurfaces, while the angular
velocity tensor is in fluid mechanics language equivalent to the
vorticity \emph{vector} defined via $\omega^i = \epsilon^{ijk} \;
\Omega_{jk}$.  The above tensors result in the unique decomposition 
of $\partial_i v_j = (\nabla \otimes {\bm v})_{ij} = D_{ij}+\Omega_{ij}$ 
into a symmetric and an antisymmetric tensor. 

The components of the Riemann curvature tensor afford the basic
symmetries $R_{[\mu\nu][\rho\lambda]}= R_{[\rho\lambda][\mu\nu]}$,
which are supplemented by $R_{[\mu\nu\rho\lambda]}= 0$ and
$R_{\mu[\nu\rho\lambda]}=0$~\cite{MTW}.  The Riemann components that
need to be calculated are thus $R_{titj}$, $R_{ijkl}$, and $R_{tijk}$,
the rest follow by the (anti-)symmetry properties.  A tedious but
straightforward computation (which follows a variant of the
Gauss--Codazzi decomposition) yields
\begin{eqnarray}
\label{E:Riemann1}
R_{ijkl} &=& D_{ik} D_{jl}-D_{il} D_{jk} \,, 
\\
R_{tijk} &=& -\partial_i \Omega_{jk}    
+v_l \left( D_{kl} D_{ij} - D_{jl}D_{ik}\right)\,,
\\
R_{titj} &=& -\partial_t D_{ij} + \left({\bm D}{\bm \Omega}
+ {\bm \Omega}{\bm D}\right)_{ij}
-\left({\bm D}^2\right)_{ij}
\nonumber\\
& &
-v_k v_{k,ij} 
+v_k v_l \left( D_{kl} D_{ij} - D_{jk}D_{il}\right)\,.
\end{eqnarray} 
Here we have defined 
$\left({\bm D}{\bm \Omega}+ {\bm \Omega}{\bm D}\right)_{ij}
\equiv D_{ik}\Omega_{kj}+ \Omega_{ik} D_{kj}$, and similarly
$\left({\bm D}^2\right)_{ij} \equiv D_{ik} D_{kj}$.

The appearance and interpretation of the Riemann components may be
greatly simplified if we consider them in an orthonormal, locally
Minkowskian tetrad frame $\{e^a{}_\mu\}$. Greek indices denote the
usual spacetime indices, Roman letters from the beginning of the
alphabet indicate tetrad indices, while Roman letters from the middle
of the alphabet denote space indices.  Whenever there is any chance of
confusion, carets on indices are used to indicate that the components
are given in the tetrad frame.  The tetrad frame $\{ e^a{}_\mu\}$ is
defined by
\be
 g_{\mu\nu} = \eta_{ab} \; e^a{}_\mu\; e^b{}_\nu.
\ee
In the simplest gauge it is given by
\begin{eqnarray}
e^{\tt}{}_t &=& 1 \,,\qquad
e^{\tt}{}_i =  0,
\nonumber
\\
e^{\ii}{}_t &=& -v^i \,,\qquad 
e^{\jj}{}_i =  \delta^{\jj}{}_i\,.
\label{tetrad}
\end{eqnarray}
The inverse basis satisfies
\be 
g^{\mu\nu} = \eta^{ab} \; e_a{}^\mu \; e_b{}^\nu.
\ee
Note the use of index placement to distinguish $e^a{}_\mu$ from its
inverse  $e_a{}^\mu$. Hence $e^a{}_\mu \; e_b{}^\mu =\delta^a{}_b$ as well as
$e_a{}^\mu\; e^a{}_\nu = \delta^\mu{}_\nu$. In a time plus space
decomposition
\begin{eqnarray}
{e}_{\tt}{}^t &=& 1 \,, \qquad
{e}_{\ii}{}^t =  0\,,
\nonumber\\
{e}_{\tt}{}^i &=& v^i \,, \qquad
{e}_{\ii}{}^j =  \delta_{\ii}{}^{j}\,.
\label{tetrad-inverse}
\end{eqnarray}
Thus, for any given vector with components 
$X^\mu$ the components in the various frames are related by
\be
X_a \equiv e_a{}^\mu \; X_\mu \equiv (X_{\tt}\,; X_\ii) = (X_t + v^j X_j \,; X_i)
\ee
and
\be
X^a \equiv e^a{}_\mu \; X^\mu \equiv (X^{\tt}\,; X^\ii) = (X^t\,; X^i  - v^i X^t).
\ee
These index conventions greatly simplify the formulae below.
Calculating the Riemann tensor in the tetrad frame gives
\begin{eqnarray}
\label{E:Riemann1F}
R_{\ii\jj\kk\lll} &=& D_{ik} D_{jl}-D_{il} D_{jk} \,, 
\\
\label{E:Riemann2F}
R_{\tt\ii\jj\kk} &=& -\partial_i \Omega_{jk}    \,,
\\\label{E:Riemann3F}
R_{\tt\ii\tt\jj} &=& -\frac{\d}{\d t} D_{ij}-\left({\bm D}^2\right)_{ij}
+ \left({\bm D}{\bm \Omega}+ {\bm \Omega}{\bm D}\right)_{ij}\,,
\end{eqnarray} 
where 
\begin{equation}
\frac{\d}{\d t} =\partial_t + {\bm v}\cdot \nabla
\end{equation}
is the usual convective derivative.  The tetrad components $R_{abcd}$
tell us how a Lagrangian observer moving with the fluid perceives the
curvature of the effective space-time described by the
\Painleve--Gullstrand metric (\ref{PGmetric3+1}).  

The components in the tetrad and co-ordinate frames are related by
\be
R_{\alpha\beta\gamma\delta} = 
e^a{}_\alpha \; e^b{}_\beta \; e^c{}_\gamma \; e^d{}_\delta \; R_{abcd}\,.
\ee
In the tetrad frame, the Ricci tensor
\begin{equation}
R_{ab}=R^c{}_{acb}= 
- R_{\tt a\tt b}+ R_{\kk a \kk b}
\end{equation}
has the components 
\begin{eqnarray}
R_{\tt\tt} 
&=& R_{\kk\tt\kk\tt}= R_{\tt\kk\tt\kk}
\nonumber\\
&=& 
-\frac{\d}{\d t} {\rm Tr}\, {\bm D} - {\rm Tr}({\bm D}^2)
\,,
\\
R_{\tt\ii} 
&=& -R_{\tt\kk\kk\ii} 
\nonumber\\
&=& \partial_k\Omega_{ki} = \frac12 \Delta v_i -\frac12 \partial_i ({\rm Tr}\, {\bm D})
\nonumber \\
&=& -\frac12 \left(\nabla\times {\bm \omega}\right)_i\,,
\\
R_{\ii\jj} &=& -R_{\tt\ii\tt\jj} + R_{\kk\ii\kk\jj}
\nonumber\\ 
&=& 
\frac{\d}{\d t} D_{ij}
-\left({\bm D}{\bm \Omega}+ {\bm \Omega}{\bm D}\right)_{ij}
+{D}_{ij}\, {\rm Tr}\,{\bm D},
\end{eqnarray}
where we remind the reader that we have defined the vorticity vector 
\begin{equation}
\omega_i = \omega^i = \epsilon^{ijk} \; \Omega_{jk}
=({\rm rot}\,{\bm v})_i = (\nabla \times {\bm v})_i \,.
\end{equation}
The curvature scalar thus becomes
\begin{eqnarray}
R &=& R_{ab} \; \eta^{ab} = -R_{\tt\tt} + R_{\kk\kk}
\nonumber\\
&=& 2 \frac{\d}{\d t} {\rm Tr}\,{\bm D} + \left({\rm Tr}\, 
{\bm D}\right)^2 
+{\rm Tr}({\bm D}^2)\,, \label{R}
\end{eqnarray}
and contains the trace of the deformation tensor and the trace of its
square, but not the vorticity.  Finally, the Einstein tensor takes the
form
\begin{eqnarray}
G_{\tt\tt} &=& R_{\tt\tt}+\frac12 R
\nonumber\\
&=& \frac12 \left({\rm Tr}\, {\bm D}\right)^2 
-\frac12 {\rm Tr}({\bm D}^2)\,,
\\
G_{\tt\ii} &=&
R_{\tt\ii} = -\frac12 \left(\nabla\times {\bm \omega}\right)_i\,,
\\
G_{\ii\jj} &=& 
R_{\ii\jj}-\frac12\delta_{\ii\jj}\; R
\nonumber\\ 
&=& 
\frac{\d}{\d t} \left( D_{ij} -\delta_{ij} {\rm Tr}\, {\bm D}\right)
+{\rm Tr}\,{\bm D}\left({D}_{ij} 
-\frac12 \delta_{ij}{\rm Tr}\,{\bm D}\right) 
\nonumber\\
& & -\frac12 \delta_{ij} {\rm Tr}({\bm D}^2)
-\left({\bm D}{\bm \Omega}+ {\bm \Omega}{\bm D}\right)_{ij}\,.
\end{eqnarray}

We emphasise that although the Ricci and Einstein tensors are
nontrivial, and certainly objects of physical interest, there is at
this level no need for or justification for imposing Einstein
equations --- though these Ricci and Einstein tensors are properties
of the flow, they are not directly related to the stress-energy tensor
generating that flow and thus the effective space-time curvature
experienced by the quasiparticles. In superfluids, for example, the
``Einstein action'' proportional to the curvature scalar (\ref{R}) is
smaller than the simple kinetic energy of the superflow by the factor
$a^2/l^2$, where $a$ is the atomic scale and $l$ the scale on which
the velocity field varies \cite{GrishaPhysicsReports}, so that the
``Einstein action'' is subdominant in determining the velocity field.

It is sometimes convenient to work with the conformally invariant,
traceless part of curvature. This is given by the Weyl tensor
\cite{HawkingEllis}:
\be
C_{abcd} = R_{abcd} + \eta_{a[d} R_{c]b}  +\eta_{b[c} R_{d]a} 
+\frac13 R \,\eta_{a [c} \eta_{d]b},
\ee
where the brackets indicate anti-symmetrization on the indices they
enclose.  This gives
\begin{eqnarray}
\label{E:Weyl1F}
C_{\ii\jj\kk\lll} &=& R_{\ii\jj\kk\lll} +\delta_{\ii[\lll}\, R_{\kk]\jj}
+\delta_{\jj[\kk}\, R_{\lll]\ii} +\frac13 R \, 
\delta_{\ii[\kk}\delta_{\lll] \jj}\,, 
\\
\label{E:Weyl2F}
C_{\tt\ii\jj\kk} &=& 
-\partial_i \Omega_{jk} - \frac12 
\delta_{i[j}(\nabla\times{\bm\omega})_{k]}   \,,
\\\label{E:Weyl3F}
C_{\tt\ii\tt\jj} &=&  
-\frac12\frac{\d}{\d t} \left( 
D_{ij} - \frac13 \delta_{ij}{\rm Tr}({\bm D}) 
\right)
\nonumber\\
&& -\left({\bm D}^2\right)_{ij} +\frac13 \delta_{ij} {\rm Tr}({\bm D}^2)
\nonumber\\
& & +\frac12  {\rm Tr}({\bm D}) \left( 
D_{ij} - \frac13\delta_{ij} ({\rm Tr}{\bm D}) 
\right)
\nonumber\\
&&
+ \frac12 \left({\bm D}{\bm \Omega}+ {\bm \Omega}{\bm D}\right)_{ij} \,.
\end{eqnarray}

%-------------------------------------------------------------------------
\section{Examples}
%-------------------------------------------------------------------------

%-------------------------------------------------------------------------
\subsection{General iso-pycnal flows}
%-------------------------------------------------------------------------

Suppose now that the flow is not only iso-tachic (constant speed of
sound) but also iso-pycnal (constant background density).  This
corresponds to an ``almost incompressible'' fluid such as water. The
major change from the previous section is the simplification that
comes from the continuity equation:
\be
{\d \rho\over \d t} = 0 
\quad \implies \quad 
\nabla \cdot {\bm v} = 0
\quad \implies \quad 
{\rm Tr}\;{\bm D} = 0.
\ee
The form of the Riemann tensor is not affected, though for the Ricci
tensor we now have
\begin{eqnarray}
R_{\tt\tt} 
&=& - {\rm Tr}({\bm D}^2)\,,
\\
R_{\tt\ii} 
&=&  \frac12 \Delta v_i\,,
\\
R_{\ii\jj} &=&
\frac{\d}{\d t} D_{ij}
-\left({\bm D}{\bm \Omega}+ {\bm \Omega}{\bm D}\right)_{ij}.
\end{eqnarray}
The Ricci scalar simplifies to
\begin{eqnarray}
R &=& {\rm Tr}({\bm D}^2)\,.
\end{eqnarray}
Thus the Ricci curvature scalar is positive semidefinite for
iso-pycnal flows, and vanishes if and only if the deformation ${\bm
D}$ is zero.

The Einstein tensor is now
\begin{eqnarray}
G_{\tt\tt} &=& 
-\frac12 {\rm Tr}({\bm D}^2)\,,
\\
G_{\tt\ii} &=&
\frac12 \Delta v_i\,,
\\
G_{\ii\jj} &=& 
\frac{\d}{\d t} D_{ij}
-\frac12 \delta_{ij} {\rm Tr}({\bm D}^2)
-\left({\bm D}{\bm \Omega}+ {\bm \Omega}{\bm D}\right)_{ij}\,.
\end{eqnarray}
Finally the Weyl tensor for iso-pycnal flows reduces to
\begin{eqnarray}
\label{E:Weyl1F-py}
C_{\ii\jj\kk\lll} &=& R_{\ii\jj\kk\lll} +\delta_{\ii[\lll}\, R_{\kk]\jj}
+\delta_{\jj[\kk}\, R_{\lll]\ii} +\frac13 R \, 
\delta_{\ii[\kk}\delta_{\lll] \jj},
\\
\label{E:Weyl2-py}
C_{\tt\ii\jj\kk} &=& -\partial_i \Omega_{jk} 
%-\partial_l \delta_{i[j}\Omega_{k]l}   \,,
+  \delta_{i[j}\Delta v_{k}] \,,
\\\label{E:Weyl3F-py}
C_{\tt\ii\tt\jj} &=&  -\frac12\frac{\d}{\d t} D_{ij}
+ \frac12 \left({\bm D}{\bm \Omega}+ {\bm \Omega}{\bm D}\right)_{ij} 
\nonumber\\
&&-\left({\bm D}^2\right)_{ij} +\frac13 \delta_{ij} {\rm Tr}({\bm D}^2)
\,.
\end{eqnarray}

%-------------------------------------------------------------------------
%\section{Examples of incompressible two-dimensional flow}
%-------------------------------------------------------------------------
\subsection{Shear flow}
%-------------------------------------------------------------------------
As a first simple example of a nontrivial incompressible flow
(Tr$\,{\bm D}=0$), consider the flow with constant shear
\be
{\bm v}= \omega_0 \, (0,x,0) \label {shearflow}
\ee
which has both constant deformation $D_{xy}=D_{yx}=\frac12 \,\omega_0$
and constant vorticity $\omega_z = \omega_0
=2\Omega_{xy}=-2\Omega_{yx}$ (all other components vanishing)
\cite{benabdallah}.  The Riemann curvature components are
\begin{eqnarray}
R_{\tt\ii\tt\jj} &=& 
-\frac14 \;\omega_0^2 \;{\cal P}_{ij},
\nonumber\\
R_{\tt\ii\jj\kk} &=& 0,
\nonumber\\
R_{\ii\jj\kk\lll} &=& 
\frac14 \;\omega_0^2 \;(\theta_{ik}\theta_{jl} -\theta_{il}\theta_{jk}),
\end{eqnarray}
where $\theta_{ik}= \theta_{ki}$ is unity if $(ik)=(xy)$ and zero
otherwise.  The projection operator \be {\cal P}_{ij}\equiv
\delta_{ij}-n_i n_j\ee where ${\bm n}=(0,0,1)$ is a unit vector in $z$
direction ensures that the curvature has nonzero components only in
the $x$ and $y$ directions.

For the Ricci and Einstein tensors
\begin{eqnarray}
R_{\tt\ii} &=& R_{\ii\jj}=0,
\nonumber\\
R_{\tt\tt} &=& -\frac12 \;\omega_0^2 = {\rm Tr} ({\bm D}^2)\,,
\nonumber\\
R &=& \frac12 \; \omega_0^2,
\nonumber\\
G_{\tt\tt} &=& -\frac14 \;\omega_0^2\,, 
\nonumber\\
\quad G_{\ii\jj} &=&  -\frac14 \; \omega_0^2 \; \delta_{ij}\,,
\nonumber\\
\quad G_{\tt\ii} &=&  0 \,.
\end{eqnarray}
Thus the quasiparticles are seen in their effective space-time to be
moving on a (3+1)-dimensional manifold of constant scalar curvature,
with radius of curvature inversely proportional to the shearing rate
$\omega_0$.

%-------------------------------------------------------------------------
\subsection{Vortex flow of constant circulation}
%-------------------------------------------------------------------------
A somewhat more interesting case is the constant-circulation flow in the
$x$-$y$ plane
\begin{equation}
v_y = \frac{\gamma x}{x^2+y^2}\,,\qquad v_x = 
-\frac{\gamma y}{x^2+y^2}
\label{vortexflow}
\end{equation}  
appropriate to a vortex flow well outside the central core, where
the circulation is $\oint {\bm v}\cdot {\bm d}s = 2\pi \gamma$. In this
case you would not want to trust the geometry for $r< r_c =
\gamma $ because at $r=r_c$ the flow goes supersonic.  This flow has
\begin{eqnarray}
D_{xx} &=& \frac{2\gamma xy}{r^4} = -D_{yy}\nonumber\\
D_{xy} &=& \frac{\gamma (y^2-x^2)}{r^4}= D_{yx}  
\nonumber\\
D_{iz} &=& D_{zi} = 0 \nonumber\\
\Omega_{ij}&=&0\,.
\end{eqnarray}
Note the ``duality'' between the vortex core and the far field. In the
core the deformation rate is zero and the vorticity is non-zero, while
in the far field it is the vorticity that is zero and deformation that
is non-zero.  The Riemann curvature tensor takes the form:
\begin{eqnarray}
R_{\xx\yy\xx\yy} &=& {\rm det}\, {\bm D} = -\frac{\gamma^2}{r^4},
\nonumber\\
R_{\tt\ii\jj\kk} &=&  0,
\nonumber\\
R_{\tt\ii\tt\jj} &=& -({\bm v}\cdot \nabla) D_{ij} - 
\left({\bm D}^2\right)_{ij}
\nonumber\\
 &=& -({\bm v}\cdot \nabla ) D_{ij}-
\frac{\gamma^2}{r^4} {\cal P}_{ij}.
\label{RiemannVortex}
\end{eqnarray}
More explicitly
\begin{eqnarray}
R_{\tt\xx\tt\xx} &=& \frac{\gamma^2}{r^6} \left(y^2-3x^2 \right)\,,
\nonumber\\
R_{\tt\yy\tt\yy} &=&  \frac{\gamma^2}{r^6} \left(x^2-3y^2 \right)\,,
\nonumber\\
R_{\tt\xx\tt\yy} &=& -\frac{4\gamma^2 xy}{r^6}\,,
\nonumber\\
R_{\tt\ii\tt\jj} &=& -\frac{\gamma^2}{r^6}\left(4 x_i x_j -\delta_{ij} \; r^2 
\right)\,.
\label{RiemannVortex2}
\end{eqnarray}
Therefore the Ricci tensor, curvature scalar, and Einstein tensor read
\begin{eqnarray}
R_{\tt\tt} &=& -\frac{2\gamma^2}{r^4} \,, \qquad 
R_{\tt\ii}= R_{\ii\jj}=0 \,,
\nonumber\\
R &=& \frac{2\gamma^2}{r^4}
\label{RicciScalarVortex}
\\
G_{\tt\tt} &=& -\frac{\gamma^2}{r^4}\,, \qquad 
G_{\ii\jj} = -\delta_{ij}\frac{\gamma^2}{r^4}\,,
\nonumber\\
G_{\tt\ii} &=& 0\,.
\end{eqnarray}
It is mildly amusing to note that the vortex geometry is uniquely
determined by the cylindrical symmetry plus the equation
$G_{ab}\propto \delta_{ab}$ (not $\eta_{ab}$).

%-------------------------------------------------------------------------
\subsection{Streaming motion past a cylinder}
%-------------------------------------------------------------------------
\label{cylindersec}
The most complex flow we discuss here is provided by the
two-dimensional streaming motion from right to left past a cylinder of
radius $a$.  According to the circle theorem \cite{milne-t}, the
complex velocity potential of such a flow is given by
\begin{equation}
w = U \left(Z+\frac {a^2} Z\right)
\end{equation}
where $Z=x+iy$ and $U$ is the velocity at infinity in negative $x$
direction.  This results in the flow
\begin{equation}
v_x =-U\left( 1+a^2\, \frac{y^2-x^2}{r^4}\right)\,,\qquad
v_y = 2Uxy\, \frac{a^2}{r^4}\,. 
\end{equation}
The velocity at infinity is restricted to be $U< 1/2$, for the maximal
velocity on the cylinder surface to be less than the speed of sound.
The formulae for deformation and vorticity (which is identically zero
for this flow) read
\begin{eqnarray}
D_{xx} &=& \frac{2U a^2 }{r^6} x \left(3y^2 -x^2 \right) = -D_{yy}\nonumber\\
D_{xy} &=& \frac{2U a^2 }{r^6} y \left(y^2 -3x^2 \right) = D_{yx}  
\nonumber\\
D_{iz} &=& D_{zi} = 0 \nonumber\\
\Omega_{ij}&=&0\,.
\end{eqnarray}
The Riemann components show that the flow past a cylinder, due to its
reduced symmetry, yields a more complicated space-time geometry for
quasiparticles than the vortex flow:
\begin{eqnarray}
R_{\xx\yy\xx\yy} &=& {\rm det}\, {\bm D} = -\frac12  {\rm Tr} ({\bm D}^2)
\nonumber\\
& = & -\frac{4U^2 a^4}{r^{6}},
\nonumber\\
R_{\tt\ii\jj\kk} &=&  0,
\nonumber\\
R_{\tt\ii\tt\jj} &=& -({\bm v}\cdot \nabla) D_{ij} - 
\left({\bm D}^2\right)_{ij}
\nonumber\\ & = & 
 -({\bm v}\cdot \nabla) D_{ij} +  {\cal P}_{ij}\,{\rm det}\, {\bm D}\,, 
\label{RiemannCylinder}
\end{eqnarray}
where the last line reads more explicitly
\begin{eqnarray}
R_{\tt\xx\tt\xx} &=& 
\frac{2U^2a^2}{r^8}\left[ a^2(y^2 -5x^2) +3(x^4 - 
6 x^2y^2 + y^4) \right],
\nonumber\\
R_{\tt\yy\tt\yy} &=&  
\frac{2U^2a^2}{r^8} \left[ a^2(x^2 -5y^2)- 3(x^4 - 
6 x^2y^2 + y^4) \right],
\nonumber\\
R_{\tt\xx\tt\yy} &=& 
-\frac{12U^2a^2}{r^8} xy \left(a^2-2x^2+2y^2\right).
\label{RiemannCylinder2}
\end{eqnarray}
These latter components show that the ``circulation'' $Ua^2$ is {not}
the only relevant parameter of the flow, in contrast to the
constant-circulation vortex case, as we may expect from the reduced
symmetry of the flow past the cylinder.

The curvature scalar
\begin{equation}
R= \frac{8U^2 a^4}{r^{6}}
\end{equation}
decays much more quickly with distance from the cylindrical object
than the curvature of the vortex flow, Eq. (\ref{RicciScalarVortex}).

%-------------------------------------------------------------------------
\subsection{Rigid rotation}
%-------------------------------------------------------------------------
The simplest example of a nontrivial incompressible flow (Tr$\,{\bm
D}=0$) is pure rotation ${\bm v}= \Omega (-y,x,0)$, which has zero
deformation $D_{ij}=0$, and constant vorticity $\omega_z = \omega_0
=2\Omega_{xy}=-2\Omega_{yx}=2\Omega$ (all other components vanishing).
This flow is appropriate for instance deep inside the core of a
vortex where the fluid effectively rotates as a ``rigid'' body.  (In
ordinary fluids this happens because viscosity dominates in the core;
in superfluids there is a more dramatic effect in that the superfluid
goes normal close enough to the core.) Also note that the core has a
maximum size given by $|{\bm v}| = 1$, that is $r_c = 2/\omega_0$.

For the rigid rotation flow it is easy to see that the Riemann
curvature tensor is identically zero, either (1) by brute force
application of the above formulae, or more subtly (2) by going to a
rotating frame (of angular velocity $\Omega = 2\omega_0$) in which the
velocity is identically zero, evaluating the Riemann tensor there
(where it is blatantly zero), and transforming back to the rotating
frame.  Although the Riemann tensor is identically zero, there is
interesting physics going on: The fact that pure rotation leads to
zero Riemann curvature is ultimately responsible for the fact that
equations (\ref{E:Riemann1}) and (\ref{E:Riemann1F}) do not contain
any terms quadratic in $\Omega$, a result that otherwise has to be
simply asserted based on explicit calculation.

Additionally, we emphasise that even though the Riemann tensor is
zero, the Christoffel symbols are definitely not zero. Indeed
\begin{eqnarray}
\Gamma^i{}_{tt} &=& - \Omega^2 r \; {\hat r}_i\,,
\\
\Gamma^i{}_{tj} &=& \Omega_{ij} = {1\over2} \epsilon_{ijk}\, \omega^k\,.
\end{eqnarray}
These two portions of the Christoffel symbols are of course simply
representing the centrifugal and Coriolis pseudo-forces. All other
components are zero.

A further (approximate) example of such a flow is encountered if one
considers the coarse-grained flow induced by a {\em lattice} of
vortices \cite{Tkachenko66}. An (infinite) lattice rotates as if it
were a solid body, with a vortex density $n_v= \Omega/\pi \gamma$
prescribed by the rotation velocity $\Omega$ and the circulation
$2\pi \gamma$, assumed to be equal for each individual vortex. 
For the vortex lattice,
it follows from the vanishing of the Riemann curvature that a
collimated quasiparticle beam can pass a (sufficiently dilute) lattice
without (on average) being deflected.

%-------------------------------------------------------------------------
\section{Geodesic deviation}
\begin{figure}[htbp]
\psfrag{s}{\LARGE $\bm s$}
\psfrag{v}{\LARGE ${\bm v}_\infty$}
\psfrag{n}{\LARGE $\bm n$}
\psfrag{u}{\LARGE $u^i {\bm e}_i$}
\vbox{
\hfil
\scalebox{0.56}{\includegraphics{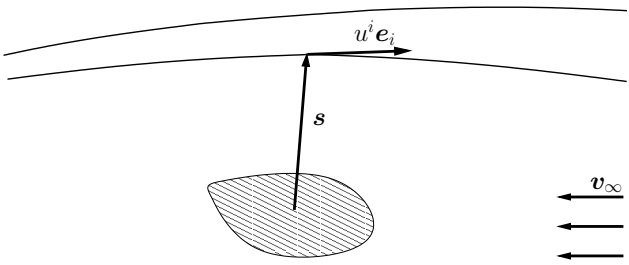}}
\hfil
}
\bigskip
\caption{
%------------------------------
The quasiparticle geodesic deviation at a distance vector $\bm s$
caused by an object placed in a flow with velocity ${\bm v}_\infty$ at
infinity (the generic case of the situation in section
\ref{cylindersec}).
%------------------------------
}
\label{Fig1}
\end{figure}

%-------------------------------------------------------------------------
An invariant measure of the strength of a flow pattern as regards its
influence on quasiparticle motion may be defined to be the value of
the curvature scalar $R \propto s^{-\kappa}$ at a certain given
distance $s$ from the flow-generating object (cf. Fig. \ref{Fig1},
illustrating the generic situation of flow past an object placed in a
homogeneous stream).  Among the flows discussed in the previous
section the shear flow is strongest in that sense (because the ``flow
generating object'' is covering all space, $\kappa=0$), followed by
the vortex flow ($\kappa= 4$) and the flow past the cylinder ($\kappa
= 6$).  Finally rigid rotation, which has zero $R$ and is
``flat'' ($\kappa =\infty$).  It is the simplest conceivable
nontrivial (\ie, inhomogeneous) flow 
with the property of having all $R_{abcd}$ equal to zero.

\smallskip

A nonvanishing Riemann tensor leads to tidal (relative) acceleration
of nearby geodesics, described by the Jacobi equation of geodesic
deviation for quasiparticles:
\begin{equation}
\frac{D^2 {n}^\alpha}{\d\lambda^2} 
+ R^\alpha{}_{\beta\gamma\delta} u^\beta n^\gamma u^\delta
=0 .
\label{E:geodev}
\end{equation}
The above relation gives the covariant relative acceleration of two
nearby geodesics, with null tangent vectors $\bm u$ separated by the
displacement vector $\bm n$, and with the geodesics affinely
parametrized by $\lambda$. (At this stage all we need to know is that
use of an ``affine parameter'' simplifies many formulae; in the
following section we will derive a relationship between the affine
parameter and physical Newtonian time $t$.)  The fact that the
constant time slices of the metric (\ref{lineelement}) are conformally
identical to flat Cartesian space in three dimensions, entails that
the space-time curvature of the quasiparticle world is reflected in a
relative acceleration of quasiparticle rays in the Newtonian lab world
of nonrelativistic hydrodynamic flow.

Consider a family of geodesics in the $x$-direction, with tangent
vector ${\bm u}^a = (u^\tt, u^\tt, 0, 0)$ and a purely spacelike
separation in the $y$-direction ${\bm n} = (0, 0, \delta y, 0)$.  We
then have
\be 
\frac{D^2 [\delta y]}{\d\lambda^2} +
\left\{ \left(R_{\yy\tt\yy\tt} + R_{\yy\xx\yy\xx} \right) (u^\tt)^2 \right\} 
[\delta y] = 0 \,.
\label{E:deviation}
\ee
This can be viewed as a parametrically driven harmonic oscillator
(driven in the affine parameter $\lambda$), with ``frequency''
\be
\Omega(\lambda) = u^\tt \; \sqrt{ R_{\yy\tt\yy\tt} + R_{\yy\xx\yy\xx} }.
\ee
Physically this means that by looking at the components of the Riemann
tensor we can see if the effective geometry \emph{locally} acts as a
focussing lens [corresponding to $\Omega(\lambda)$ real] or as a
diverging lens [corresponding to $\Omega(\lambda)$ imaginary]. Since
(in the focussing case, and assuming a reasonably uniform medium) two
initially parallel geodesics will focus down to a point after an
elapse of affine parameter $\delta \lambda = \pi /\Omega(\lambda) $,
the corresponding local focal length is (in physical distance units)
given by
\be
f^{\rm local} 
= {\pm \pi\over  \sqrt{ ||R_{\tt\yy\tt\yy} + R_{\xx\yy\xx\yy}|| } }.
\label{Localf}
\ee
Note the strengths and weaknesses of this concept --- it provides a
{\emph{local}} position and orientation dependent notion of focal
length appropriate for nearly parallel geodesics (nearly parallel
quasiparticles; so one is automatically working ``on axis'' and
ignoring ``spherical abberation''), but this definition of $f^{\rm
local}$ does in general not provide significant global information.
If the Riemann tensor is strongly inhomogeneous, varying on length
scales significantly smaller than $f^{\rm local}$, then this concept
of local focal length is not particularly useful.  In particular, in
the vortex geometry of~\cite{Vortex}, with flow (\ref{vortexflow}),
the focussing effect we had in mind was a global effect due to
quasiparticles passing by opposite sides of the vortex core, with
impact parameter $b$ --- this is not a situation that can be described
by the Jacobi equation.  The global result obtained there for $f =
f^{\rm global}= ({2b^3/ 3\pi r_c^2})
\left[1+O\left({r_c/b}\right)\right]$, is {\em not} the local $f^{\rm
local}$ defined above. Indeed two initially parallel quasiparticles
passing by on the same side of the vortex core will be driven apart
from each other by geodesic deviation --- it is this effect that leads
to the ``cylindrical abberation'' of the lens discussed
in~\cite{Vortex}.

A case where the local focal length {\em does} acquire global meaning
is the shear flow (\ref{shearflow}), for which the focal length
(\ref{Localf}) becomes a constant:
\be 
f^{\rm shear} %=f^{\rm local}
=  \frac{\sqrt{2}\pi}{\omega_0}. 
\ee  
The focal length is in this case bounded by the atomic length scale
itself, simply due to the requirement that the concept of
hydrodynamics makes sense.  This further strengthens the notion of the
shear flow being the strongest possible flow as regards its influence
on quasiparticle motion, because any other flow has more stringent
bounds on the {\em global} $f$.

One useful refinement of the local focal length concept introduced in
equation (\ref{Localf}) is to consider null geodesics (quasiparticle
paths) propagating in an arbitrary unit direction $\bar{\bm u}$ and
then use indices $M$ and $N$ to denote the two spatial directions
perpendicular to $\bar{\bm u}$. Then the local focal length can be
generalized to a $2\times2$ matrix
\be
f^{\rm local}_{MN} = 
{\pm \pi\over  
\sqrt{ ||R_{\tt M \tt N} + R_{\ii M\jj N} \; {\bar u}^\ii \; {\bar u^\jj} || } }.
\ee
The square root and inverse is to be taken in the matrix sense, and
the two eigenvalues of $f_{MN}$ are the two principal focal lengths
along the direction $\bar{\bm u}$. If these eigenvalues differ it is a
signal of astigmatism.

%-------------------------------------------------------------------------
\section{Non-affine parameterization of null geodesics}
%-------------------------------------------------------------------------

While the use of affine parameters for null geodesics is standard in
general relativity, it should be borne in mind that in the present
\Painleve--Gullstrand context there is a preferred temporal foliation
provided by the Newtonian time parameter $t$. It is worth the
technical bother of using the non-affine parameterization in terms of
$t$ here in order to make aspects of the physics clearer.

In general, we know that along any null geodesic there will be some
relationship between affine parameter $\lambda$ and Newtonian time
$t$. For instance we can assert
\begin{equation}
\d \lambda = \exp[\zeta(t)] \; \d t. \label{nonaffine}
\end{equation}
In the affine parameterization the geodesic equation for a null curve
is just
\[
u^\mu \nabla_\mu u^\nu = 0;   \qquad u^\nu \equiv {\d x^\nu \over\d\lambda}.
\]
If we choose a non-affine parameterization
\[
\bar u^\mu \nabla_\mu \bar u^\nu = \dot \zeta(t) \; \bar u^\mu;   
\qquad 
\bar u^\nu \equiv {\d x^\nu \over\d t}.
\]
The geodesic equation becomes
\be
{\d^2 x^\mu\over\d t^2} = 
- \Gamma^\mu{}_{\alpha\beta} \; {\d x^\alpha\over\d t} \;{\d x^\beta\over\d t} 
+ \dot \zeta(t) \; {\d x^\mu\over\d t}. 
\ee
In this form it is clear that the physical acceleration of the
quasiparticle is related to gradients in the \Painleve--Gullstrand
metric.  It is extremely useful to derive an explicit relationship
between the affine parameter $\lambda$ and the physical Newtonian time
$t$. To do this let's start with the notion of a stationary geometry
(technically: there exists a timelike Killing vector; colloquially: a
time-independent geometry). The timelike Killing vector takes the form
\be
K^\mu = (1; \; \vec 0); \qquad K_\mu = \left(-[1-{\bm v}^2]; \; -{\bm v} \right).
\ee
The tangent vector to the null geodesic is denoted
\be
u^\mu = {\d x^\mu\over \d\lambda} = 
{\d t\over \d\lambda} \; \left(1; \; {\d {\bm x} \over \d t} \right).
\ee
It is a standard theorem that the 3+1 inner product between a geodesic
tangent vector and a Killing vector is conserved, as long as the
geodesic is affinely parameterized. Thus
\be
g_{\mu\nu} \; K^\mu \; u^\nu = 
{\d t\over \d\lambda} \;
\left[ 1 - {\bm v}^2 + {\bm v} \cdot  {\d {\bm x} \over \d t} \right] 
= \hbox{constant}.
\ee
On the other hand, because $u^\mu$ is a null vector
\be
1 - {\bm v}^2 + 2  {\bm v} \cdot  {\d {\bm x} \over \d t} - 
\left|{\d {\bm x} \over \d t}\right|^2 = 0.
\ee
Eliminating between these two equations, we can normalize in such a way
that
\be
{\d t\over \d\lambda}  = \exp[-\zeta(t)] = 
\left[ 1 - {\bm v}^2 + \left({\d {\bm x} \over \d t}\right)^2 \right]^{-1}.
\ee
That is
\be
\zeta(t) = 
\ln\left[ 1 - {\bm v}^2 + \left({\d {\bm x} \over \d t}\right)^2 \right].
\ee
If the fluid is not moving, then ${\bm v}=0$ and $|{\d {\bm x} /\d
t}|=1$ so $t\propto\lambda$. If the fluid is moving we simply have to
live with this position-dependent factor relating the affine parameter
$\lambda$ (in terms of which the geodesic equations are most easily
written down) to the Newtonian time parameter $t$ (in terms of which
the physical acceleration is most easily calculated).

In a similar manner, the Jacobi equation can be rewritten as
\begin{equation}
\frac{D^2 {n}^\alpha}{\d t^2} - \dot \zeta(t) \; \frac{D {n}^\alpha}{\d t}
+ R^\alpha{}_{\beta\gamma\delta} \bar u^\beta n^\gamma \bar u^\delta
=0.
\label{E:geodev2}
\end{equation}
While this looks somewhat messier than the affinely parameterized
Jacobi equation (\ref{E:geodev}), the physics is the same.  In
particular if we start with two initially parallel null geodesics
($Dn/\d t =0$ at $t=0$), and assume a locally homogeneous medium, we
are led to the same notion of local focal length as discussed in the
previous section.

%-------------------------------------------------------------------------
\section{Discussion}
%-------------------------------------------------------------------------
We have shown how the generation of curved Riemannian space-time
geometries for quasiparticles is possible based purely on the velocity
pattern of a nonrelativistic flow.  Conversely, one might conceive of
solving for a flow field from a given space-time geometry. This is a
highly nonlinear problem, as becomes obvious from the relations
(\ref{E:Riemann1F})--(\ref{E:Riemann3F}). It is, however, certainly no
more nonlinear or complicated than solving the Einstein equations of
general relativity themselves.  While the \Painleve-Gullstrand
geometry discussed here does not provide us with the most
generic case (remember that the constant time surfaces are
(conformally) flat; for generalizations allowing for more general
space-time metrics see \cite{CSM}), it shows that the underlying
kinematical structure of a curved space-time can in principle be
perfectly nonrelativistic. The dynamical identification of this
effective geometry with general relativity, \ie, imposing the Einstein
equations, is a more advanced step \cite{GrishaPhysicsReports}, but is
possible in principle as well.

There are several generalizations of the current analysis that would
be of interest: (1) If the quasiparticle propagation speed ($c$, local
speed with respect to the background medium) is varying then the
geometry exhibits ``index gradient'' effects in addition to effects
generated by the motion of the medium. While technically
straightforward, the relevant calculations of the Riemann tensor are
computationally messy and the physical interpretation is not so clear
(unless the medium is completely at rest; in which case one recovers
standard ``index gradient'' physics). (2) If the density varies from
place to place, then it is necessary to distinguish the ``geometrical
quasiparticle'' regime (the analogue of geometrical optics) from the
``wave quasiparticle regime'' (the analogue of wave optics). In the
geometrical approximation the results of the present paper can be carried
over; in the wave regime one needs to carry out an analysis in terms
of Green functions and wave equations; the entire armoury of
quasiparticle trajectories as null geodesics of the effective metric
breaks down and must be replaced by a more fundamental wave
description.

In summary: The use of pseudo--Riemannian geometry has important
applications well beyond the confines of general relativity. In
particular quasiparticle propagation in condensed matter systems can
often be characterized in terms of an ``effective'' spacetime
geometry; most easily described in \Painleve--Gullstrand form. If the
background medium is a fluid, then the Riemann curvature (and
Christoffel symbols, \etc) can be calculated in terms of shear
(deformation) and vorticity of the fluid.  Ultimately this analysis
relates the focussing and deflection of quasiparticles to the 
properties of the fluid flow.
%-------------------------------------------------------------------------
\section*{Acknowledgments}
%-------------------------------------------------------------------------
URF acknowledges support by the {\em Deutsche Forschungsgemeinschaft}
(FI 690/2-1) and the ESF Programme ``Cosmology in the Laboratory''.
MV was supported by the US Department of Energy.
%------------------------------------------------------------------------

%------------------------------------------------------------------------
\end{document}